\documentclass[11pt]{article} 
\usepackage{mystyle-new}
\usepackage{authblk}
\usepackage[T1]{fontenc}
\usepackage{epsfig,amsmath} 
\usepackage{hepnames,hepunits}
\usepackage{hyperref}
\usepackage{color}
\usepackage{graphicx}
\definecolor{red}{rgb}{1,0,0}
\def\lesssim{\ \hbox{\raise 2pt \hbox{$<$} \kern -13pt
                     \lower 3pt \hbox{$\sim$}}\ }
\def\greatersim{\ \hbox{\raise 2pt \hbox{$>$} \kern -13pt
                     \lower 3pt \hbox{$\sim$}}\ }

\def\lsim{\mathrel{\rlap{\lower4pt\hbox{\hskip1pt$\sim$}}
    \raise1pt\hbox{$<$}}}                % less than or approx. symbol
\def\gsim{\mathrel{\rlap{\lower4pt\hbox{\hskip1pt$\sim$}}
    \raise1pt\hbox{$>$}}}                % greater than or approx. symbol

\input epsf.tex
\def\desepsf(#1 width #2){\epsfxsize=#2 \epsfbox{#1}}

\newenvironment{tolerant}[1]{\par\tolerance=#1\relax}{ \par }
\usepackage{amsmath,bm}
\usepackage{lineno}
\usepackage{cite,mcite}
\usepackage{tikz}
\usepackage[symbol]{footmisc}

\providecommand{\DOI}[1]{\href{http://dx.doi.org/#1}}

\usepackage{wrapfig}
\usepackage{orcidlink}

\begin{document}

\title{
A Science4Peace initiative:\\
Alleviating the consequences of sanctions in international scientific cooperation}
\author[1]{
A.~Ali~\orcidlink{0000-0002-1939-1545}, 
M.~Barone~\orcidlink{0000-0002-2115-4055}, 
S.~Brentjes~\orcidlink{0000-0002-8205-8550}, 
D.~Britzger~\orcidlink{0000-0002-9246-7366}
M.~Dittmar~\orcidlink{0009-0008-0472-6251}, 
T.~Ekel\"of~\orcidlink{0000-0002-7341-9115}, 
J.~Ellis~\orcidlink{0000-0002-7399-0813},
S.~Fonseca~de~Souza~\orcidlink{0000-0001-7830-0837}, 
A.~Glazov~\orcidlink{0000-0002-8553-7338}, 
A.V.~Gritsan~\orcidlink{0000-0002-3545-7970}, 
R.~Hoffmann~\orcidlink{0000-0001-5369-6046},
H.~Jung \thanks{Corresponding author and contact person: hannesjung@science4peace.com}~\orcidlink{0000-0002-2964-9845}, 
M.~Klein~\orcidlink{0000-0002-8527-964X}, 
V.~Klyukhin~\orcidlink{0000-0002-8577-6531}, 
V.~Korbel~\orcidlink{0009-0007-9781-9344}, 
P.~Kokkas~\orcidlink{0009-0009-3752-6253}, 
P.~Kostka~\orcidlink{0000-0002-8551-3272}, 
U.~Langenegger~\orcidlink{0000-0001-6711-940X}, 
J.~List~\orcidlink{0000-0002-0626-3093}, 
N.~Raicevic~\orcidlink{0000-0002-2386-2290}, 
A.~Rostovtsev~\orcidlink{0000-0001-9906-0764}, 
A.~Sabio~Vera~\orcidlink{0000-0003-0228-5313}, 
M.~Spiro~\orcidlink{0009-0007-1113-1056}, 
G.~Tonelli~\orcidlink{0000-0003-2606-9156}, 
P.~van~Mechelen~\orcidlink{0000-0002-8731-9051}, 
J.~Vigen~\orcidlink{0000-0002-2050-7701}
}

\begin{titlepage} 
\maketitle
%\vspace*{-11cm}
%\begin{flushright}
%version 1.5 \\
%%\today
%\end{flushright}
%\vspace*{+12cm}

\begin{abstract}
The armed invasion of Ukraine by the Russian Federation has adversely affected the relations between Russia and Western countries. Among other aspects, it has put scientific cooperation and collaboration into question and changed the scientific landscape significantly. 
Cooperation between some Western institutions and their Russian and Belarusian partners were put on hold  after February 24, 2022. The CERN Council decided at its meeting in December 2023 to terminate cooperation agreements with Russia and Belarus that date  back a decade.

CERN is an international institution with UN observer status, and has so far played a role in international cooperation which was independent of national political strategies. 

We argue that the Science4Peace idea still has a great value and scientific collaboration between scientists must continue, since fundamental science is by its nature an international discipline. A ban of  scientists participating in  international cooperation and collaboration is against the traditions, requirements and understanding of science.

We call for measures to reactivate the peaceful cooperation of individual scientists on fundamental research in order to stimulate international cooperation for a more peaceful world in the future. Specifically, we plead for finding ways to continue this cooperation through international 
organizations, such as CERN  and JINR.
\end{abstract} 
\end{titlepage}

\section{The historical international cooperation at CERN  and the Science for Peace mission} 
\label{Intro}
In the aftermath of World War II, nations came together and formed the United Nations (UN) with the purpose, as stated in the first article of the UN charter~\cite{UN-charter}, "... to take effective collective measures for the prevention and removal of threats to the peace". With more than 100 wars and military conflicts since then~\cite{TodaysWars}, we are further away than ever from this ideal, marking a significant failure of diplomacy to prevent those wars.

\begin{wrapfigure}{r}{0.45\textwidth}
\begin{center}
\vskip -1.5cm
\includegraphics[scale=.24,angle=-90]{ 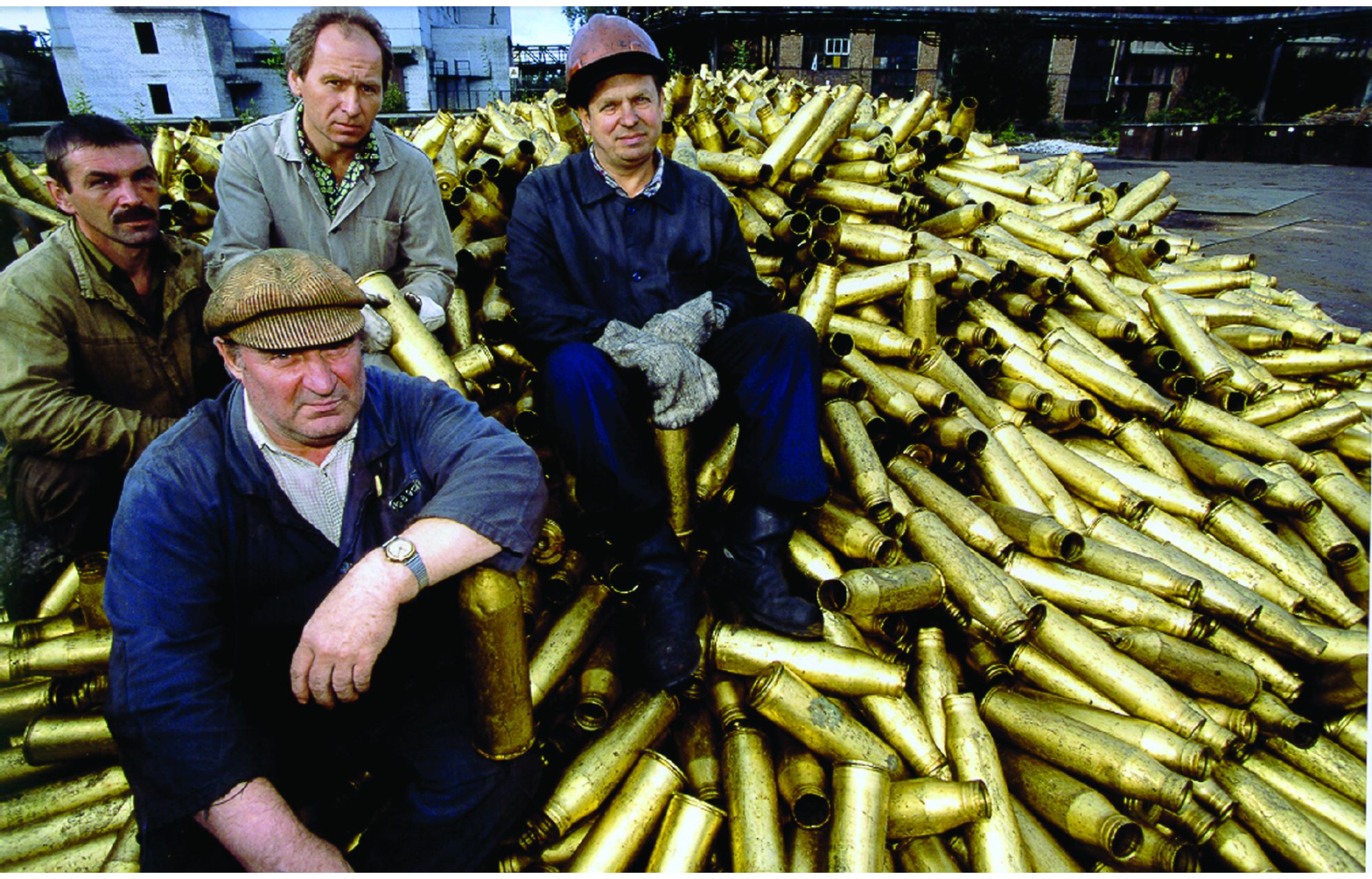}
\caption{Thousands of brass shells in Russian military storage were melted-down for use in the CMS detector (from~\protect\cite{CMS-and-Russian-Navy-shells})}
\label{CMS_hcal}
\end{center}
\end{wrapfigure}

In a similar spirit as the UN, CERN was founded in 1954 to bring nations together through peaceful scientific collaboration. Remarkably, just one year after its foundation, cooperation between CERN and Soviet scientists began via the Joint Institute for Nuclear Research (JINR) in Dubna \cite{Lock:1975fz} and then, in 1967, via the Institute for High Energy Physics in Protvino. In 2014, on the occasion of CERN's 60th anniversary, the former Director-General Rolf Heuer wrote that "CERN has more than fulfilled the hopes and dreams of advancing science for peace"~\cite{Heuer-S4P}. 

The construction of the Large Hadron Collider (LHC)~\cite{Evans:2008zzb,LHC-CERN} at CERN as well as the experimental detectors of the big LHC experiments was possible because of significant contributions from the Russian and Belarusian institutes. In particular, a part of the calorimeter of the CMS detector~\cite{CMS-and-Russian-Navy-shells} was built from melted-down brass Russian navy shells (Fig.~\ref{CMS_hcal}), a wonderful expression of the spirit of the {\it Swords to Ploughshares} sculpture at the UN headquarters.

CERN is the international center for particle physics, with the world's largest particle collider, the LHC, and hosting the largest international scientific  collaborations ATLAS, CMS, ALICE and LHCb of up to 4000 scientists each. CERN is the only place at present, where  fundamental physics at the frontier of the  highest energies can be performed. It is  where the Future Circular Collider (FCC)~\cite{FCC-StudyGroup,FCC:2018byv} is being discussed, a new project that, if approved, is planned to start operations in the 2040's and expected to deliver data until the end of this century.

CERN was established first at an intergovernmental meeting of UNESCO in Paris in December 1951 \cite{cern-history} and has been granted UN - Observer status~\cite{CERN-UNobserver} and has therefore responsibilities beyond those of national institutes that are controlled by national policies.

CERN has served as a model for the SESAME project~\cite{SESAME-unesco,SESAME-home} in the Middle East, as well as for the proposal to build  a similar scientific infrastructure in the Western Balkans called SEEIST~\cite{SEEIST-home}, bringing together scientists from Albania, Kosovo, Bosnia and Herzegovina, Montenegro and Serbia.

Fundamental research, being funded by public resources for the advancement of knowledge, is not just global public good but is also powerful instruments for intercultural dialogue and peace  -- especially during times of crisis. One of CERN's basic principles is that results in fundamental research  {\it shall have no concern with work for military requirements and the results of its experimental and theoretical work shall be published or otherwise made generally available}, as written in CERN's convention~\cite{CERN-convention}.  Several other institutes and universities have also declared, that their research is only for non-military purposes in a the so-called {\it Civil Clause}~\cite{civil-clause-Japan,civil-clause}.

\section{The change in science policy and the damage to international relations}

The armed invasion of Ukraine by the Russian Federation at the end of February 2022 and the suffering inflicted on countless innocent civilians, including scientists, has received strong international condemnation. 
Despite pro-war statements from the heads of some Russian organisations, many Russian physicists opposed the war and immediately signed petitions against it~\cite{OpenLetterRussianScientistsAgainstWar}. In March 2022, as a reaction to the war in Ukraine, many national Western science institutions put bans on their historical scientific cooperation with Russian institutions. However, in an article in the CERN courier in 2022 the former CERN director Herwig Schopper has argued "Science for Peace? More than ever"~\cite{CERNcourier-Schopper}. 

The International Union of Pure and Applied Physics (IUPAP)~\cite{IUPAP-association}) has taken a clear position against exclusion of scientists  from participating in conferences or events on the basis of their nationality or their affiliation~\cite{IUPAP-affiliation,IUPAP-statemen-on-war}. The president of IUPAP found very clear words against exclusion of scientists at a panel discussion reported in Ref.~\cite{Albrecht:2023xxi}.

In February 2023 the LHC experiments at CERN decided to  remove all affiliations from the authors from the Russian and Belarusian institutions in publications\footnote{The original documents of the decisions of the experiments are not available publicly, only internally.} \cite{LHCdecisionOnAffilation} (examples in Refs.~\cite{ALICE:2023csm,ATLAS:2023gzn,CMS:2023xpx,LHCb:2024ier}). This was a discriminatory act based on the country of affiliation. In contrast, other non-CERN international collaborations continued with their original author-list, listing all affiliations on equal footing (see e.g. Refs.~\cite{Muong-2:2023cdq,Belle-II:2024xzm,DayaBay:2024xye}).
It is crucial to note that international scientific cooperation with Russia still continues elsewhere, e.g.  XFEL~\cite{XFEL-partner}, ESA~\cite{ESA-cooperation}, ITER~\cite{ITER-cooperation}, and ISS~\cite{ISS-cooperation} and at the Japanese  research center KEK and the Belle II experiment that it hosts. {The LHC Collaborations could have adopted a similar model of cooperation}, according to which scientific collaboration is not to be interpreted as an endorsement of the policies of the governments in which countries the institutes are located. This would be a  more equitable compromise, enabling continued scientific collaboration despite differences in the political positions, for which scientists bear no responsibility.

The ban on long established scientific cooperation unexpectedly also affected CERN, whose Council -- where the member states of CERN are represented -- deliberated on the status of the existing cooperation agreements with Russian and Belarusian Institutes -- and decided to terminate them~\cite{CERNcouncilICABelarusTermiantion,CERNcouncilICARussiaTermiantion}. 
However, the consequences of this decision went much further than removing the framework for establishing scientific collaboration, it also called for termination of existing agreements, therewith denying access of Russian and Belarusian scientists to scientific data and equipment that they jointly own (with just one of many examples shown in Fig.~\ref{CMS_hcal}).

The decision of the CERN Council in December 2023 to terminate cooperation with Russia and Belarus marks a significant change in science diplomacy: this decision breaks with CERN's mission  of {\it Science for Peace} \cite{CERNmission}. 
The consequences of the decision of the CERN Council  can hardly be overestimated as it may  affect future international projects and it raises the following troublesome questions: will countries still invest a significant amount of financial and personal resources in projects, where they risk being excluded at some stage~? Will countries in the Middle- and Far East, from Africa or South-America still have trust in CERN ? Or will they  invest in  projects in other regions? Or  will there be more investment in military -  instead of fundamental - research ?

The decision of the CERN Council to terminate the cooperation agreements  could lead to a break in the cooperation between European and Russian science in general and can lead to irreversible consequences on an international scale.  Several countries may begin to question their cooperation with CERN. 

Cooperations and collaborations are to a large extent based on trust, trust that the investment will pay off and trust that agreements to cooperate will be respected. All this is now under question.  
It has created mis-trust, a shock and frustration that the scientific community as a whole did not oppose clearly the discriminatory political decisions to terminate participation in research at CERN.

Our Russian and Belarusian colleagues have suddenly become {\it personae non gratae} at CERN. Some of the consequences of this exclusion were already summarized in FAQ's published by the CERN users office~\cite{CERN_user_office_RU-BY_institutes}, immediately after the decision of the CERN Council in December 2023.

As already stated, limiting international scientific collaboration is against the advancement of knowledge, which is not just a global public good but also a powerful instrument for intercultural dialogue and peace -- especially during times of crisis. If we take the UN charter seriously, we must support measures that help to prevent and remove threats to  peace.  
It is important to note, that the UN (with CERN holding UN observer status~\cite{CERN-UNobserver}) did not endorse any scientific exclusion of researchers from any international cooperation. 
The LHC has been a prime example for friendly cooperation on a global scale, naturally including Ukrainian scientists who now deserve our special support.

Excluding a significant part of the scientific community from international projects, like the LHC~\cite{LHC-CERN} at CERN puts politics before science.   It intensifies tensions with Russian and Belarusian colleagues who oppose the Russian invasion and is against the basic principles on which CERN was founded. The adoption of sanctions can be used as a template in future conflicts. However, as in the United Nations, we, the international science community, must insist that especially in difficult times cooperation  in international organizations must continue, rather than countries being expelled from such organizations.

Excluding a whole community from international projects like the LHC means, that they are excluded from analysis of data that they have helped to record using detectors that they have developed, and hence lose credit for any forthcoming discoveries. Moreover, those scientists are excluded from shaping fundamental science at the forefront via international interactions  and discussion that  are essential ingredients in  peaceful cooperation between people, nations and states in the present and the future.

 This is all the more regrettable because throughout  its 70-year history CERN has been  a role-model for collaborative scientific work and international collaboration, and other important international projects like SESAME~\cite{SESAME-home} and SEEIST~\cite{SEEIST-home} were designed having the success of CERN in mind. If CERN is to keep this  role, also for future projects and collaborative efforts,  it needs to operate as a model for a World Laboratory, where all those interested in common scientific goals and shared responsibilities are welcome. Shutting the doors for countries, with whom CERN member states have political differences, would seriously compromise its model character.

In a  recent publication~\cite{Albrecht:2023xxi}, the enormous consequences of sanctions in science were discussed, and it was argued, that they are very harmful for scientific progress and scientific culture. 
In an opinion article {\it Science needs cooperation, not exclusion} in the CERN courier of March 2024 \cite{opinion-vie-CERN-courier} arguments for a continuing dialogue across all borders are given. {\it The Geneva Observer} reported  on the consequences of the  CERN Council decision in~\cite{geneva-observer-2024}.

\section{The Science4Peace Initiative}

In order to keep a certain level of trust and responsibility in an international organization, everything must be done to ensure that scientists from Russia and Belarus who have contributed with know-how, research, building parts of the detector, responsibilities in experimental analyses and in physics research will continue to be able to use for non-military scientific purposes any data and knowledge resulting from the experiments  until their completion.

We call for a return to an equitable, non-discriminatory treatment of all authors who have contributed to scientific results. A straight-forward solution has been adopted by the Belle II collaboration, who waived all affiliations in scientific publications~\cite{BelleII-Statement}. 

We therefore propose, as immediate action,  to limit negative consequences in the present situation through the following steps:
\begin{itemize}
\item Grant continued access to  data and any knowledge resulting from  experiments to the collaborating scientists without any discrimination. In the present crisis, CERN should work out  a {\it modus operandi} by fostering collaborations through international institutes, such as JINR, Dubna, Russia, so as to allow collaborating scientists continuing access to CERN;
\item Sign scientific publications in a fair manner, either only with names (omitting the names of affiliated institutes and laboratories), 
or else state their affiliations, on an equal basis for all,
acknowledging also the support received from the organizations and funding agencies for carrying out the experiments.
\end{itemize}

Each individual scientist should be able to decide which topic to work on and who to collaborate with.  These decisions are covered by the generally accepted principle of {\it Freedom of Science}, which has constitutional or legal status in most EU Member States~\cite{BonnDeclarationOnFreedomOfSecientificResearch} and many other countries and is covered by the {\it International Covenant on Economic, Social and Cultural Rights} by the United Nations~\cite{UN-InternationalCovenant}.
Therefore it is only appropriate that the scientists themselves play a leading role in the scientific planning and organization of their research.

Each individual scientist believing in the universal and international ideas of scientific research and in the basic ideas of {\it Science for Peace}, shall be supported to contribute by starting new and dedicated collaborations with scientists who are otherwise excluded.  New projects and cooperations are rather easy to execute in theory and phenomenology, and are being continued today.  In experimental particle physics, the situation is more difficult, as access to detectors and accelerators as well as to the data which are recorded, is needed. However, for several years an Open Data Portal~\cite{CERN-OpenData} has existed, where the LHC experiments provide a subset of their recorded data together with the relevant software and tools for further analysis. Some publications based on these Open Data have already been performed (e.g., in Refs~\cite{Larkoski:2017aa,Tripathee:2017aa}). This may serve as an interim, partial template for open access to LHC data and they may help to not loose or establish contacts, especially for those that are currently being excluded from the collaborations. 
\\

{\it Going beyond these steps, we propose, as a Science4Peace Initiative}:
\begin{itemize}
\item Allowing and encouraging international scientific cooperation among all countries committed to the United Nations charter;
\item Continuing scientific communication between individuals and continuing to produce common scientific publications on fundamental physics,  without political constraints;
\item Initiating dedicated projects in theory and phenomenology, as well as in experimental physics based on openly accessible resources, for  interested scientists on the basis of universal non-military scientific goals, independent of the nationality, gender or color of the participating scientists;
\item Allowing wide conference participation, independently of restrictions, by holding conferences at least in hybrid mode, which will also reduce significantly flights and the associated ecological footprint~\cite{Gorlinger:2023aa};
\item Organizing international summerschools for students that are either fully online or open to students from anywhere, with full coverage provided for travel and local expenses.
\end{itemize}

The enormous consequences resulting from the decision of the CERN Council  
 affect  not only the present ongoing research but, even more importantly, the future of basic scientific research, and career prospects of young scientists. Therefore this decision demands common, far-sighted and cooperative actions in  reply. We encourage the international scientific community to support this, as a Science for Peace initiative, with more ideas and initiatives to counter the threatened scientific schism.

\vskip 1 cm 
\begin{tolerant}{8000} 
\noindent 
{\bf Acknowledgments.} 
We thank all participants of the {\it Science4Peace Forum} for supporting discussions. In particular, we thank Vladimir Lipp for helpful discussions and constructive comments on the text.

\noindent 

\end{tolerant} 
\vskip 0.6cm 

\raggedright  
\providecommand{\href}[2]{#2}\begingroup\raggedright\endgroup

%\bibliography{/Users/jung/Bib/hannes-bib}

\begin{thebibliography}{10}%
\makeatletter
\providecommand{\hrefCMSnoop }[0]{\@secondoftwo}%
\makeatother
\providecommand{\doi}{\texttt{doi:}\begingroup \urlstyle{tt}\Url}

\bibitem{UN-charter}
\href {https://www.un.org/en/about-us/un-charter/full-text}{``United Nations
  Charter'',} 26 June, 1945.

\bibitem{TodaysWars}
\href
  {https://geneva-academy.ch/galleries/today-s-armed-conflicts}{\mbox{Geneva
  Academy (Academy of International Humanitarian Law and Human Rights)},
  ``Today's Armed Conflicts'',} 2023.

\bibitem{CMS-and-Russian-Navy-shells}
\href {https://cms.cern/book/export/html/1202}{\mbox{CMS Collaboration}, ``The
  CMS HCAL and Russian Navy Shells'',} Nov, 2011.

\bibitem{Lock:1975fz}
\href {https://cds.cern.ch/record/186009/files/CERN-75-07.pdf?version=1}{W.~O.
  Lock, ``{A History of the Collaboration Between the European Organization for
  Nuclear Research (CERN) and the Joint Institute for Nuclear Research (JINR),
  and with Soviet Research Institutes in the USSR, 1955-1970}''.} CERN Yellow
  Report 75-7, 1975.

\bibitem{Heuer-S4P}
\href {https://home.cern/news/opinion/cern/celebration-science-peace}{R.~Heuer,
  ``A celebration of science for peace''.} CERN, 2014.

\bibitem{Evans:2008zzb}
\hrefCMSnoop {}{``{LHC Machine}'',} \textit{ JINST} \textbf{ 3} (2008) S08001.

\bibitem{LHC-CERN}
\href {https://home.cern/science/accelerators/large-hadron-collider}{``The
  Large Hadron Collider (LHC)''.}

\bibitem{FCC-StudyGroup}
\href {https://fcc.web.cern.ch/}{``The Future Circular Collider''.}

\bibitem{FCC:2018byv}
\href
  {https://www.interactions.org/press-release/international-collaboration-publishes-concept-design-post}{{FCC}
  Collaboration, ``{FCC Physics Opportunities}: {Future Circular Collider
  Conceptual Design Report Volume 1}'',} \textit{ Eur. Phys. J. C} \textbf{ 79}
  (2019) 474.

\bibitem{cern-history}
\href {https://home.cern/about/who-we-are/our-history}{CERN, ``CERN History''.}

\bibitem{CERN-UNobserver}
\href
  {https://home.cern/news/press-release/cern/cern-granted-status-observer-united-nations-general-assembly}{CERN,
  ``CERN is granted the status of observer to the United Nations General
  Assembly'',} 2012.

\bibitem{SESAME-unesco}
\href
  {https://en.unesco.org/courier/2018-4/sesame-scientific-excellence-middle-east}{``SESAME:
  scientific excellence in the Middle East''.}

\bibitem{SESAME-home}
\href {https://www.sesame.org.jo/}{``Synchrotron-light for Experimental Science
  and Applications in the Middle East (SESAME)''.}

\bibitem{SEEIST-home}
\href {https://seeiist.eu/}{``The South East European International Institute
  for Sustainable Technologies (SEEIIST)''.}

\bibitem{CERN-convention}
\href {https://cds.cern.ch/record/330625?ln=de}{CERN, ``Convention for the
  establishment of a European organization for nuclear research''.} CERN
  document server, 1953.

\bibitem{civil-clause-Japan}
\href
  {https://www.scj.go.jp/ja/info/kohyo/pdf/kohyo-23-s243-en.pdf}{\mbox{Science
  Council of Japan}, ``Statement on Research for Military Security'',} 2017.

\bibitem{civil-clause}
\href {https://en.wikipedia.org/wiki/Civil_clause}{Wikipedia, ``Civil
  Clause''.} Website, last edited 27 October 2023.

\bibitem{OpenLetterRussianScientistsAgainstWar}
\href {https://www.t-invariant.org/2022/02/we-are-against-war-en/}{``Open
  letter of Russian scientists and science journalists against the war with
  Ukraine'',} Feb, 2022.

\bibitem{CERNcourier-Schopper}
\href
  {https://cerncourier.com/a/science-for-peace-more-than-ever/}{H.~Schopper,
  ``Science For Peace? More than ever!''.} CERN Courier, 2022.

\bibitem{IUPAP-association}
\href {https://iupap.org/documents/statutes-bylaws/articles}{IUPAP, ``Articles
  of Association'',} 2021.

\bibitem{IUPAP-affiliation}
\href
  {https://iupap.org/wp-content/uploads/2022/07/use_iupap_affiliation_at_conferences.pdf}{IUPAP,
  ``Use of IUPAP Affiliation at Conferences'',} 2022.

\bibitem{IUPAP-statemen-on-war}
\href
  {https://iupap.org/2022/03/01/iupap-statement-on-the-events-occurring-in-ukraine2022-03-01/}{IUPAP,
  ``IUPAP statement on the events occurring in Ukraine'',} 2022.

\bibitem{Albrecht:2023xxi}
\hrefCMSnoop {}{M.~Albrecht {et~al.}, ``{Beyond a Year of Sanctions in
  Science}'',}
\newblock 2023.
\newblock
  \href{http://www.arXiv.org/abs/2311.02141}{\texttt{arXiv:2311.02141}}.

\bibitem{LHCdecisionOnAffilation}
\href {https://www.nature.com/articles/d41586-023-00503-5}{R.~V. Noorden, ``LHC
  physicists resolve stalemate over Russian authors''.} Nature, 2023.

\bibitem{ALICE:2023csm}
\hrefCMSnoop {}{{ALICE} Collaboration, ``{Measurements of long-range
  two-particle correlation over a wide pseudorapidity range in p\textendash{}Pb
  collisions at $ \sqrt{s_{\textrm{NN}}} $ = 5.02 TeV}'',} \textit{ JHEP}
  \textbf{ 01} (2024) 199,
  \href{http://www.arXiv.org/abs/2310.07490}{\texttt{arXiv:2310.07490}}.

\bibitem{ATLAS:2023gzn}
\hrefCMSnoop {}{{ATLAS} Collaboration, ``{Studies of new Higgs boson
  interactions through nonresonant HH production in the $ b\overline{b}\gamma
  \gamma $ final state in pp collisions at $ \sqrt{s} $ = 13 TeV with the ATLAS
  detector}'',} \textit{ JHEP} \textbf{ 01} (2024) 066,
  \href{http://www.arXiv.org/abs/2310.12301}{\texttt{arXiv:2310.12301}}.

\bibitem{CMS:2023xpx}
\hrefCMSnoop {}{{CMS} Collaboration, ``{Search for new Higgs bosons via
  same-sign top quark pair production in association with a jet in
  proton-proton collisions at $\sqrt{s}$ = 13 TeV}'',} \textit{ Phys. Lett. B}
  \textbf{ 850} (2024) 138478,
  \href{http://www.arXiv.org/abs/2311.03261}{\texttt{arXiv:2311.03261}}.

\bibitem{LHCb:2024ier}
\hrefCMSnoop {}{{LHCb} Collaboration, ``{Measurement of the Branching Fraction
  of $B^{0} \rightarrow J/\psi \pi^{0}$ Decays}'',}
  \href{http://www.arXiv.org/abs/2402.05528}{\texttt{arXiv:2402.05528}}.

\bibitem{Muong-2:2023cdq}
\hrefCMSnoop {}{{Muon g-2} Collaboration, ``{Measurement of the Positive Muon
  Anomalous Magnetic Moment to 0.20~ppm}'',} \textit{ Phys. Rev. Lett.}
  \textbf{ 131} (2023) 161802,
  \href{http://www.arXiv.org/abs/2308.06230}{\texttt{arXiv:2308.06230}}.

\bibitem{Belle-II:2024xzm}
\hrefCMSnoop {}{{Belle-II} Collaboration, ``{Measurement of $CP$ asymmetries in
  $B^0\to\eta'K^0_s$ decays at Belle II}'',}
  \href{http://www.arXiv.org/abs/2402.03713}{\texttt{arXiv:2402.03713}}.

\bibitem{DayaBay:2024xye}
\hrefCMSnoop {}{{Daya Bay} Collaboration, ``{First measurement of the yield of
  $^8$He isotopes produced in liquid scintillator by cosmic-ray muons at Daya
  Bay}'',}
  \href{http://www.arXiv.org/abs/2402.05383}{\texttt{arXiv:2402.05383}}.

\bibitem{XFEL-partner}
\href {https://www.xfel.eu/organization/partner_countries/index_eng.html}{XFEL,
  ``Partner countries of XFEL'',} 2024.

\bibitem{ESA-cooperation}
\href
  {https://www.esa.int/Enabling_Support/Space_Transportation/International_cooperation}{ESA,
  ``International Cooperation - European Space Agency (ESA)''.}

\bibitem{ITER-cooperation}
\href {https://www.iter.org/proj/Countries}{ITER, ``ITER Members'',} 2024.

\bibitem{ISS-cooperation}
\href
  {https://www.nasa.gov/international-space-station/space-station-international-cooperation/}{ISS,
  ``International Space Station (ISS) -- International Cooperation''.}

\bibitem{CERNcouncilICABelarusTermiantion}
\href
  {https://council.web.cern.ch/sites/default/files/c-e-3758-Resolution%20BY%2015%20December%202023.pdf}{\mbox{CERN
  council}, ``Termination of the International Cooperation Agreement between
  CERN and the Republic of Belarus'',} 2023.

\bibitem{CERNcouncilICARussiaTermiantion}
\href
  {https://council.web.cern.ch/sites/default/files/c-e-3757-Resolution%20RU%2015%20December%202023.pdf}{\mbox{CERN
  council}, ``Termination of the International Cooperation Agreement between
  CERN and the Russian Federation'',} 2023.

\bibitem{CERNmission}
\href {https://home.cern/about/who-we-are/our-history}{CERN, ``Science for
  Peace''.}

\bibitem{CERN_user_office_RU-BY_institutes}
\href
  {https://usersoffice.web.cern.ch/sites/default/files/pdf/Russia%20Ukraine/FAQ-for%20those%20affiliated%20to%20RU-BY%20institutes_18.12.2023_v.3.pdf}{\mbox{CERN
  user office}, ``FAQ for those affiliated to RU-BY institutes'',} 2023.

\bibitem{opinion-vie-CERN-courier}
\href
  {https://cerncourier.com/a/science-needs-cooperation-not-exclusion/}{H.~Jung,
  ``Science needs cooperation, not exclusion''.} CERN Courier, 2024.

\bibitem{geneva-observer-2024}
\href
  {https://www.thegenevaobserver.com/at-cern-when-politics-collides-with-science-international-scientific-collaboration-survived-the-cold-war-and-other-conflicts-russias-war-on-ukraine-might-have-killed-it/}{P.~Mottaz,
  ``Exclusive: The Inside Story of How CERN Sanctioned Russia''.} The Geneva
  Observer, 2024.

\bibitem{BelleII-Statement}
\href
  {https://science4peace.com/ewExternalFiles/Belle%20II%20Response%20to%20the%20Ongoing%20Russian%20Invasion%20of%20Ukraine-Revised13Oct2022-endorsed20Oct2022.pdf}{\mbox{Belle
  II Collaboration}, ``Belle II Response to the Ongoing Russian Invasion of
  Ukraine''.}

\bibitem{BonnDeclarationOnFreedomOfSecientificResearch}
\href
  {https://www.bmbf.de/bmbf/shareddocs/downloads/files/_drp-efr-bonner_erklaerung_en_with-signatures_maerz_2021.pdf?__blob=publicationFile&v=1}{\mbox{Ministerial
  Conference on the European Research Area}, ``Bonn Declaration on Freedom of
  Scientific Research'',} 2020.

\bibitem{UN-InternationalCovenant}
\href
  {https://treaties.un.org/doc/treaties/1976/01/19760103%2009-57%20pm/ch_iv_03.pdf}{\mbox{United
  Nations}, ``International Covenant on Economic, Social and Cultural
  Rights'',} 1967.

\bibitem{CERN-OpenData}
\href {https://opendata.cern.ch/}{``CERN Open Data Portal''.}

\bibitem{Larkoski:2017aa}
A.~Larkoski\href {https://arxiv.org/abs/1704.05066}{ {et~al.}, ``Exposing the
  QCD Splitting Function with CMS Open Data'',}
  \href{http://www.arXiv.org/abs/1704.05066}{\texttt{arXiv:1704.05066}}.

\bibitem{Tripathee:2017aa}
A.~Tripathee\href {https://arxiv.org/abs/1704.05842}{ {et~al.}, ``Jet
  Substructure Studies with CMS Open Data'',}
  \href{http://www.arXiv.org/abs/1704.05842}{\texttt{arXiv:1704.05842}}.

\bibitem{Gorlinger:2023aa}
\href {https://doi.org/10.1038/s44168-023-00069-y}{S.~G{\"o}rlinger, C.~Merrem,
  M.~Jungmann, and N.~Aeschbach, ``An evidence-based approach to accelerate
  flight reduction in academia'',} \textit{ npj Climate Action} \textbf{ 2}
  (2023) 41.

\end{thebibliography}

\end{document}